# Foldscope: Origami-based paper microscope

Short Title: Origami-based paper microscope

Classification: Physical Sciences, Engineering

James Cybulski[1], James Clements[2], Manu Prakash[2]

[1]Department of Mechanical Engineering,

[2]Department of Bioengineering, Stanford University

318 Campus Drive, Stanford, CA 94305

Corresponding Author:

Prof. Manu Prakash

Stanford University

318 Campus Drive, Stanford, CA

617-820-4811

manup@stanford.edu

# Title: Foldscope: Origami-based paper microscope

**Authors:** James Cybulski[1], James Clements[2], Manu Prakash[2]*

**Affiliations:**

[1]Department of Mechanical Engineering, Stanford University

[2]Department of Bioengineering, Stanford University

*Correspondence to: manup@stanford.edu

**Abstract**: Here we describe an ultra-low-cost origami-based approach for large-scale manufacturing of microscopes, specifically demonstrating brightfield, darkfield, and fluorescence microscopes. Merging principles of optical design with origami enables high-volume fabrication of microscopes from 2D media. Flexure mechanisms created via folding enable a flat compact design. Structural loops in folded paper provide kinematic constraints as a means for passive self-alignment. This light, rugged instrument can survive harsh field conditions while providing a diversity of imaging capabilities, thus serving wide-ranging applications for cost-effective, portable microscopes in science and education.

**Significance Statement:**
Combining the principles of origami with optical design, we present ultra-low cost brightfield, darkfield, and fluorescence microscopes designed for rugged applications in science and education.

Microscopes are ubiquitous tools in science, providing an essential, visual connection between the familiar macro-world and the remarkable underlying micro-world. Since the invention of the microscope, the field has evolved to provide numerous imaging modalities with resolution approaching 250nm and smaller (1). However, some applications demand non-conventional solutions due to contextual challenges and tradeoffs between cost and performance. For example, *in situ* examination of specimens in the field provides important opportunities for ecological studies, biological research, and medical screening. Further, ultra-low cost DIY microscopes provide means for hands-on science education in schools and universities. Finally, this platform could empower a worldwide community of amateur microscopists to capture and share images of a broad range of specimens.

Cost-effective and scalable manufacturing is an integral part of "frugal science and engineering" (2). For example, manufacturing via folding has emerged as a powerful and general-purpose design strategy with applications from nanoscale self-assembly (3) to large-aperture space telescopes (4). More recently, possibilities of folding completely functional robots have been explored (5-7), with actuators, sensors and flexures integrated in a seamless fashion. Modern micro-lens fabrication technology is another prime example of scalable manufacturing. Although the use of high-curvature miniature lenses traces back to Antony van Leeuwenhoek's seminal discovery of microbial life forms (8), manufacturing micro-lenses in bulk was not possible until recently. Modern techniques such as micro-scale plastic molding and centerless ball-grinding have grown to serve numerous applications, including telecommunication fiber couplers, cell phone cameras, and medical endoscopes.

By combining principles of optical design with origami (9-11), here we present a novel platform for the fabrication of flat microscopes cheaply in bulk (figure 1). The Foldscope is an origami-based optical microscope that can be assembled from a flat sheet of paper in under 10 minutes (see supplementary video 1). Although it costs less than a dollar in parts (see Bill of Materials in table 1), it can provide over 2,000X magnification with submicron resolution, weighs less than two nickels (8.8 g), is small enough to fit in a pocket ($70 \times 20 \times 2$ mm$^3$), requires no external power, and can survive being dropped from a 3-story building or stepped on by a person (see figure 1G and supplementary video 2). Its minimalistic, scalable design is inherently application-specific instead of general-purpose, providing less functionality at dramatically reduced cost. Using this platform, we present our innovations for various imaging modalities (brightfield, darkfield, fluorescence, lens-array) and scalable manufacturing strategies (capillary encapsulation lens mounting, carrier tape lens mounting, self-alignment of micro-optics by folding, paper microscope slide).

The Foldscope is operated by inserting a sample mounted on a microscope slide (figure 1B), turning on the LED (figure 1C), and viewing the sample while panning and focusing with one's thumbs. The sample is viewed by holding the Foldscope with both hands and placing one's eye close enough to the micro-lens so one's eyebrow is touching the paper (figure 1F). Panning is achieved by placing one's thumbs on opposite ends of the top stage (colored yellow in figure 1A-C) and moving them in unison, thus translating both optics and illumination stages while keeping the stages aligned (figure 1B). Focusing is achieved using the same positioning of one's thumbs, except the thumbs are pulled apart (or pushed together). This causes tension (or compression) along the optics stage, resulting in -Z (or +Z) deflection of the micro-lens due to flexure of the supporting structure of the sample-mounting stage (figure 1C). Unlike traditional microscopes, the Foldscope anchors the sample at a fixed location while the optics and illumination stages are moved in sync.

**Design Platform**
*Construction from flat media.* The Foldscope is comprised of three stages cut from paper ─ illumination, sample-mounting, and optics ─ and assembled via folding (figure 1A-C, supplementary video 1). Other primary components include a spherical ball lens (or other micro-lenses), lens-holder apertures, an LED with diffuser or condenser lens, a battery, and an electrical switch (see figure 1D). The three stages are weaved together to form an assembled Foldscope (figure 1B,C,E) with the following features: fully-constrained X-Y panning over a $20 \times 20$ mm$^2$ region (figure 1B), flexure-based focusing via Z-travel of the optics stage relative to the sample-mounting stage (figure 1C), and a vernier scale for measuring travel distances across the sample slide with 0.5mm resolution. The total optical path length from the light source to the last lens surface is about 2.7mm (figure S1), only 1% that of a conventional microscope. Flat polymeric sheets and filters can also be inserted into the optical path, including diffusion filters for improving illumination uniformity, Fresnel lenses as condensers for concentrating illumination intensity, color filters for fluorescence imaging, and linear polarizers for polarization imaging.

*Alignment by Folding.* Folding provides a passive alignment mechanism that is used here to align the micro-lens with the light source. A sharp crease in a thin sheet of inextensible material, such as paper, of thickness $h$ introduces elastic energy of bending of the order $\sim h^3$ (12). Thus, a fold introduces buckling at the inner edge, giving variation in the exact location of the hinge and resulting in random alignment error of the order $\sim h$. To minimize this error, we introduce folding features that form a closed structural loop between the optics stage and the illumination stage. This improves alignment repeatability through elastic averaging within kinematic constraints (figure 1A; 13). We characterized alignment accuracy and repeatability by constructing twenty independent Foldscopes out of 350μm thick black cardstock and manually folding and unfolding them twenty times each (see Materials and Methods section), while measuring absolute X-Y alignment (figure S2). Assembly repeatability was assessed as the mean value of twice the standard deviation for each Foldscope (65μm in X and 25μm in Y), while assembly accuracy was assessed as the mean value of all trials (59μm in X and 67μm in Y). A higher skew in X-axis repeatability results from structurally distinct constraints implemented for the X- and Y-axes. The small assembly accuracy errors (less than 20% of the paper thickness) in both directions are consequences of the design which can be compensated by feature shifts in future designs.

*Micro-Optics and Illumination.* The Foldscope design accommodates different optical configurations, including spherical ball lenses, spherical micro-lens doublets (such as a Wollaston doublet), and more complex assemblies of aspheric micro-lenses. While more optical elements generally provide reduced aberration and improved field of view, spherical ball lenses have distinct advantages for high-volume manufacturing, including reduced part count and simplified assembly due to rotational symmetry (14-16). Since magnification varies inversely with ball-lens diameter, commonly available ball lenses provide an ample range of magnifications (under 100X to over 2,000X, as seen in table 2). The back focal length of these lenses varies drastically, thus motivating alternative lens-mounting schemes (above the optics stage, as in figure 1C, or below) and requiring samples with no coverslip for lenses with less than approximately 140μm back focal length. Equally important for image quality, the illumination source (LED plus diffuser and/or condenser lens) should provide even illumination over the field, ample intensity, narrow intensity profile, and high CRI (color rendering index). The LED used in the Foldscope consumes only 6mW of electrical power and can operate over 50 hours on a CR2032 button cell battery (figure S3A). Precise control over the illumination profile is required for high-quality microscopy (17), so integration of a condenser lens is crucial for optimal imaging (figure S3C). For low-magnification imaging applications not requiring optimal imaging, the illumination source can be removed and the Foldscope can be operated while facing an external light source.

**Design Innovations**
*High-Resolution Brightfield Microscopy.* For some applications, extending the resolution limit of the Foldscope to submicron length scales is a practical necessity. For this reason, the resolution of the single-ball-lens Foldscope was further optimized and empirically characterized. The analytical optimization was carried out for a single field point at the optical axis to assess the best achievable resolution (see Modeling and Characterization and table 2). A 1,450X Brightfield Foldscope with the configuration depicted in figure 2E was used to capture the image in figure 2A, empirically confirming submicron resolution. As shown in figure S4A, spherical ball lenses have significant wavefront error at the edge of the field defined by the aperture (aperture shown in figure S1). As a result, not all regions can be simultaneously in focus within this field. The center portion of the field, with wavefront error less than 1/5 wavenumber and low curvature and distortion, is denoted the "optimal field of view" (figure S4A-C). Thus, the best achievable resolution is attained at the expense of a reduced field of view. When a digital sensor is used in place of the naked eye, the lens fixture effectively reduces the field of view to roughly the optimal field of view.

*Fluorescence.* Conventional fluorescence microscopy typically requires an expensive illumination source for high-intensity broad-spectrum illumination and multiple optical elements with precisely defined spectral profiles. The simplified configuration of the Fluorescence Foldscope uses a high-intensity colored LED of narrow spectral width and polymeric sheets inserted in the optical path for a shortpass excitation filter and a longpass emission filter (figure 2D). A blue LED light source and commonly available gel filters (with spectral transmissivities plotted in figure S3B) were used to image 2μm diameter red poly-fluorescent polystyrene beads as shown in figure 3C. For fluorescent imaging requiring higher contrast, small pieces (3mm square or smaller) of interference filters can be used in place of the polymeric sheets at reasonable cost due to the small size.

*Lens-array and Multi-modality.* Since a micro-lens has a very small footprint, multiple optical paths can be independently configured in a single Foldscope (figure 2C,G). Such a lens array may be comprised of identical lenses or of different lenses with different magnifications and/or back focal lengths. This provides for the design of a lens-array Foldscope with several key features. For non-contiguous samples such as blood smears, a larger field of view can be obtained by overlapping a number of small fields of view. Alternatively, an optimal array pitch will give tangential non-overlapping fields of view (as seen in figure 2C), thus reducing the time required to scan a slide for a feature such as a parasite. Since individual lenses have independent optical paths, the novel capability of building multi-modality lens-array microscopes arises. One such combination is a two-by-one array of brightfield and fluorescence modalities, which could be used to scan a sample for the presence of fluorescence markers and then identify the non-fluorescent surrounding structures.

*Darkfield.* The Darkfield Foldscope configuration, shown in figure 2H, requires a diffuser, a darkfield condenser aperture (inset in figure 2D), and a condenser lens. The diffuser helps to evenly distribute light from the small LED over the aperture area, while the condenser focuses a hollow cone of light onto the specimen. Since the specimen must be placed at the focal point of the condenser, the slide thickness has to match the back focal length of the condenser plus the spacing from the condenser to the slide. A 140X Darkfield Foldscope was used to image 6μm polystyrene microspheres as shown in figure 2D.

*Capillary Encapsulation Lens Mounting.* The process of precisely mounting micro-optics to an aperture crucially governs lens performance. Therefore, a capillary encapsulation process was developed to automatically mount a ball lens while forming a circular aperture of precisely tunable diameter (see figure 3A). By partially engulfing ball lenses in an opaque polymer held between two glass substrates coated with flat nonstick PDMS (see top left of figure 3A), a precise aperture is self-assembled around the ball-lens (see Materials and Methods section for details). Pressure applied between the substrates is used to precisely tune aperture size, with greater pressure providing a larger aperture. The epoxy encapsulated lens is then adhesively mounted to a paper aperture and inserted in the Foldscope (see bottom right of figure 3A).

*Carrier Tape Lens Mounting.* Black polystyrene carrier tape is commonly used for low-cost reel-to-reel packaging of electronic components. This pre-existing infrastructure was leveraged to create low-cost mounting structures for ball lens with optimized apertures. As depicted in figure 3B, the custom thermoformed pocket holds the lens in place with a press fit, and a punched hole in the bottom of the pocket precisely defines the aperture. A single lens is cut from the carrier tape and adhesively attached to a paper aperture which is then inserted into the optics stage of the Foldscope, analogous to that shown for the epoxy encapsulated lens in figure 3A.

*Paper microscope slide*. A low-cost microscope slide was constructed out of 18mil polystyrene synthetic paper and transparent tape as shown in figure 3C. If tape is placed on only one side (either upper or lower), specimens are conveniently mounted on the exposed sticky surface of the tape. Using both upper and lower pieces of transparent tape creates a cavity for mounting wet specimens such as live algae suspended in water. Since the paper microscope slide is less than half the thickness of a standard glass slide, a spacer is required to elevate the sample closer to the lens. This is achieved by inserting the specimen slide together with two blank paper slides beneath it. Once the specimen has been viewed, the transparent tape can be removed from the synthetic paper and replaced so that the slide can be reused for many specimens. Note that the specimen depicted in figure 2D was mounted on a paper microscope slide.

**Modeling and Characterization**

*Theory and analysis.* For a brightfield Foldscope, basic measures of optical performance can be described in terms of the ball radius (r), index of refraction (n), aperture radius (a), and incident wavelength ($\lambda$; see Supplementary Material). Assuming the paraxial approximation, these include effective focal length (EFL), back focal length (BFL), and magnification (MAG). For a 300µm sapphire ball lens: EFL=172µm, BFL=22µm, and MAG=1,450X. Thus substantial magnification can be obtained, but the sample must be separated from the lens by only a fraction of the thickness of a human hair. Three additional optical performance metrics include field of view radius (FOV), numerical aperture (NA), and depth of field (DOF). These depend on aperture radius (a), the optimization of which is discussed below. For the previous example, the normalized optimal aperture radius is nOAR=a/r=0.51, giving: FOV=88µm, NA=0.44, DOF=2.8µm (see table 2).

The aperture radius controls the balance between diffraction effects from the edges of the aperture with spherical aberration effects from the lens. Therefore, a complete analytical model was created to predict the normalized optimal aperture radius (nOAR) and optimal resolution (RES), as well as the aberration coefficient (s) for a ball lens (see Supplementary Materials and table 2), yielding:

$$\text{nOAR} = (k_1/r)(\lambda/|s|)^{1/4}, \quad k_1 = (3\sqrt{10}/4\pi)^{1/4} \cong 0.9321$$
$$\text{RES} = k_2 f(\lambda/|s|)^{1/4}, \quad k_2 = 1.22(\pi/12)^{1/4}(e/10)^{1/8} \cong 0.7415$$
$$s = 1(n-1)[n + (2-n)(2n-1)]/(2rn)^3$$

The expressions for nOAR and RES are depicted as 2D design plots as a function of desired MAG in figure 4A,B and as 3D plots over n and r in figure 4C,D. For the example discussed earlier, the values for normalized aperture radius and resolution are found by locating the intersection of the lines for r=150µm and n=1.77 in the design plots. This gives nOAR=0.51 and RES=0.86µm, and corresponds to MAG=1,450X. Note that the regions enclosed by the curves in the 2D design plots represent the available design space for nOAR, RES, and MAG as defined by the range of possible values for n and r. The design curves thus make it a simple exercise to pick optimal design parameters within the space of interest. Also, one can see from figure 4B that the lower limit for the best achievable resolution in ball lenses appears to be near 0.5µm, based on the range of parameters identified for this figure.

*Numerical Modeling.* A ray-tracing numerical model was developed for the Foldscope using Zemax software to confirm the results of the analytical model and to evaluate points across the field of view (see Materials and Methods section for details). The results for nOAR and RES show very good agreement, with correlation coefficients of $R^2=0.985$ for nOAR and $R^2=0.998$ for RES (see figure 4C). The numerical modeling results across the field of view are shown in figure S4A, where the optimal field of view used for calculating the MTF (Modulus of the Optical Transfer Function) is defined. The MTF for this system is plotted in figure 4D for a field point on the optical axis and another at the edge of the optimal field of view. The field point at the optical axis shows near-diffraction-limited response, and the tangential and sagittal curves for the edge of the field drop to half of their low-frequency value at a spatial frequency of about 300 cycles/mm.

**Discussion and Conclusion**
By removing cost barriers, Foldscope provides new opportunities for a vast user base in both science education and field work for science and medicine. Many children around the world have never used a microscope, even in developed countries like the United States. A universal program providing "a microscope for every child" could foster deep interest in science at an early age. While people have known for decades that hands-on examination and inquiry is crucial in STEM (Science, Technology, Engineering, and Mathematics) education (18-19), the challenge posed by J. M. Bower to engage "all teachers and all children" (20) requires large-scale adoption of practices and broad availability of tools that were previously cost-prohibitive (21). Moreover, the opportunity to make microscopes both approachable and accessible can inspire children to examine the rich bio-diversity on our planet as amateur microscopists and to make discoveries of their own, as already seen in the field of amateur astronomy (22; see images taken by novice user with self-made Foldscope in figure 5H-J).

Disease-specific Foldscope designs are an important vision for future development (23-24). Figure 5 depicts early bench-test data, including high-magnification brightfield images of *Giardia lamblia*, *Leishmania donovani, Trypanosoma cruzi (*Chagas parasite*)*, *Escherichia coli,* and *Bacillus cereus* (figure 5A-F), and low-magnification brightfield images of *Schistosoma haematobium* and *Dirofilaria immitis* (figure 5G). Note that these include magnifications ranging from 140X to 2,180X, none of which require immersion oil. In the future, darkfield and fluorescence Foldscopes will also be adapted for diagnostics, and sensitivity and specificity will be measured for various disease-specific Foldscopes in the field as clinical validations against existing diagnostic standards.

Constructing instruments from 2D media provides other unique advantages and opportunities. Embedding flat rare-earth magnets in paper provides means for magnetic self-alignment, allowing the Foldscope to be reversibly coupled to a conventional smartphone for image capture. By printing text and images on the paper, Foldscope provides an efficient information-delivery platform for specific staining protocols, pathogen identification guides, or language-free folding instructions (figure S6). Some applications in highly infectious diseases may benefit from a disposable microscope ─ or "use-and-throw" microscopy ─ where the entire microscope can be incinerated. Also, in place of a glass slide, the 2D media also allows direct addition of the sample to a paper-based micro-fluidic assay (25) for automated staining and/or pathogen-concentration, thus yielding an independent fully-functional diagnostic system.

Future work will build upon the key features of this platform. Roll-to-roll processing of flat components and automated "print-and-fold" assembly make yearly outputs of a billion units attainable. Ongoing work with advanced micro-optics and illumination design ― including spherical GRIN lenses (26-27), aspheric multi-lens optics, and condenser lens provisions for Köhler illumination ― is expected to improve both resolution and field of view at low cost. International field-work in both diagnostics and education will provide vital inputs for further improvements. Our long-term vision is to universalize frugal science, using this platform to bring microscopy to the masses.

**Materials and Methods.**
*Ball Lenses.* The ball lenses used in constructing Foldscopes included material types borosilicate, BK7 borosilicate, sapphire, ruby, and S-LAH79. The vendors included Swiss Jewel Co, Edmund Optics, and Winsted Precision Ball. Part numbers for some select lenses include: 300μm sapphire lens from Swiss Jewel Co. (Model B0.30S), 200μm sapphire lenses from Swiss Jewel Co. (Model B0.20S), 2.4mm borosilicate lenses from Winsted Precision Ball (P/N 3200940F1ZZ00A0), 300μm BK7 borosilicate lenses from Swiss Jewel Co. (Model BK7-0.30S), and 1.0mm BK7 borosilicate lenses from Swiss Jewel Co. (Model BK7-1.00). Note that half-ball lenses from both Edmund Optics and Swiss Jewel Co. were also tested for use as condenser lenses for the LEDs.

*2D Media and Filters.* The 2D media used in constructing Foldscopes included black 105 lb card stock (ColorMates Smooth & Silky Black Ice Dust Card Stock, purchased from thePapermillstore.com), polypropylene (PressSense Durapro CC 10mil), and others. Foldscope parts were cut from 2D media using a $CO_2$ laser (Epilog Elite, Mini24). Copper tape was used for providing connectivity (by soldering) between the LED, battery, and switch. The filters used in constructing Foldscopes included Roscolux colored gel filters (including Primary Blue #80 and Fire Red #19, which approximate an Acridine Orange filter set), Roscolux diffuser filters (Tough Rolex #111), and polymeric linear polarizers (Edmund Optics P/N 86181). Each type of filter is assembled to the Foldscope by cutting out a 3-5mm square piece and adhesively attaching it to the appropriate stage with single-sided or double-stick Scotch tape. Paper microscope slides were constructed from polypropylene sheets (PressSense Durapro CC 18mil) and transparent scotch tape.

*LEDs, Switches and Power Sources.* The LEDs used in constructing Foldscopes included the Avago HSMW-CL25 (now replaced by P/N Avago ASMT CW40) white LED for brightfield Foldscopes, the Kingbright APTD1608QBC/D blue LED for fluorescence Foldscopes. The electrical slider switch was purchased from AliExpress.com ("Off/On MINI SMD Switch" from Product ID: 665019103). The power sources included Duracell 3V CR2032 button cells, Sanyo 3V CR2016 button cells (Sanyo CR2016-TT1B #8565 from Batteriesandbutter.com), and a GW Instek DC power supply (Model GPD-3303D). Button cells were used with no resistors for Foldscopes.

*Aperture Manufacturing.* This method produces inexpensive apertures through polymer encapsulation of ball lenses while preserving the optical quality of the lens and allowing multiple lenses to be encapsulated at once. The experimental setup shown in the top left of figure 3A was used to encapsulate 300μm sapphire ball lenses with aperture diameters ranging from 100μm to 214μm. The lens was sandwiched between parallel substrates (glass or silicon) coated with planar films of PDMS with thickness greater than 1mm (formed from Dow Corning Sylgard 184 PDMS). A micrometer stage was used to precisely apply pressure between the substrates to adjust the diameter of the resulting elastic deformation of the PDMS film. This diameter was measured in situ using phase contrast microscopy to set the target value for the aperture. A fast-curing opaque polymer (Smooth-On Smooth-Cast Onyx Fast Polyurethane) was then injected into the cavity and allowed to cure. Reflected light microscopy was used to measure the dimensions of the final aperture formed. Once removed from the non-stick PDMS films, the encapsulated lens was attached to the underside of the optics stage of a Foldscope

*Characterization of Self-Alignment by Folding.* Twenty independent microscopes were cut out of black 105 lb cardstock, each marked with a cross-hair in both the optics and illumination stages (see figure S2C). After folding, alignment was measured using a dissection microscope (Olympus upright, 30X magnification) via digitizing the cross-hair images, drawing lines through the center of each cross-hair (X and Y cross-hairs on both stages), and digitally measuring the X and Y displacements to characterize the alignment. Every Foldscope was iteratively folded, imaged to record X-Y alignment, and unfolded twenty times. The data was then used to assess accuracy and repeatability (see figure S2A,B).

*Sample Preparation.* Thin-blood smears of *Plasmodium falciparum* (ring stage), *Trypanosoma cruzi*, *Giardia lamblia*, *Leishmania donovani*, and *Dirofilaria immitis* were freshly prepared from cultures provided by Center for Discovery and Innovation in Parasitic Diseases (CDIPD) at UCSF. The samples were fixed in methanol and stained in freshly prepared Giemsa solution (Sigma Aldrich, #48900-500ML-F) using standard protocols before imaging. Once fixed, the slides could be used for several weeks. Bacterial samples of *Bacillus cereus* and *Escherichia coli* were provided by KC Huang Lab at Stanford University. The samples were heat fixed onto glass slides using standard procedures and gram stained using standard protocols (Fisher Scientific Gram Stain Set, Catalog No. 23-255-959). Plasmodium-infected red blood cells were taken from cultures provided by the Center for Discovery and Innovation in Parasitic Diseases (CDIPD). *Schistosoma haematobium* were provided by the Michael Hsieh Lab at Stanford University. Insects used for imaging were caught on Stanford campus and imaged after fixing in formaldehyde without any stain. No human samples were utilized in the current work.

*Image-Capture Protocol.* Brightfield images were taken using a Canon EOS 5D Mark II with the Foldscope placed 3cm away from the 100mm focal length lens and using the following settings: F/3.2, 1/30 sec. exposure, ISO-2000. An initial image was first captured using automatic white-balance and then used as a reference white balance image during data collection. Fluorescence images were taken in a similar fashion to the bright field images with typical camera settings: F/2.8, 15 sec. exposure, ISO-1000. Although not presented, images were also obtained by coupling the Foldscope to cell-phones including an iPhone using a magnetic coupler.

*Numerical Model.* Zemax software was used to model the Foldscope optics to assess optimal aperture radius and resolution. The basic model of the system consists of a ball lens, an aperture, an object at infinity, and an image plane (see figure S5A). This model requires two parameters to be independently optimized — lens-image distance and aperture radius. The analysis is carried out in four steps: 1) optimize lens-image distance in model by minimizing focusing metric (figure S5C); 2) determine search space for aperture radius as defined by empirically chosen limits on Strehl Ratio, 0.75-0.98; 3) optimize aperture radius using resolution metric (figure S5D); and 4) use Matlab surface-fitting tool to fit data for optical performance parameters as functions $F(n, r, \lambda)$ and compare with analytical model.

**References:**

1. E. Keller, R. Goldman. Light Microscopy, 8 (Cold Spring Harbor Laboratory Press, Woodbury, NY 2006).

2. G.M.Whitesides. The Frugal Way, The Economist - The World in 2012, (2011).

3. P. W. Rothemund. Folding DNA to create nanoscale shapes and patterns. Nature, 440, 297-302 (2006).

4. R. A. Hyde. Eyeglass, a large-aperture space telescope. Appl. Opt. 38 (19), 4198-4212 (1999).

5. A. M. Hoover, Ronald S. Fearing, Fast scale prototyping for folded millirobots. 2008 IEEE International Conference on Robotics and Automation 5, 886-892 (2008).

6. J. P. Whitney, P. S. Sreetharan, K. Y. Ma, R. J. Wood, Pop-up book MEMS. J. Micromech. Microeng. 21, 115021 (2011).

7. P. S. Sreetharan, J. P. Whitney, M. D. Strauss, R. J. Wood, Monolithic fabrication of millimeter-scale machines. J. Micromech. Microeng. 22, 055027 (2012).

8. J. R. Porter, Antony van Leeuwenhoek: tercentenary of his discovery of bacteria, Bacteriol. Rev. 40, 260-269 (1976).

9. M. McArthur, R. Lang. Folding paper: The infinite possibilities of Origami (International Arts & Artists, Washington, DC 2012).

10. E. D. Demaine, J. O'Rourke. Geometric Folding Algorithms: Linkages, Origami, Polyhedra (Cambridge Univ. Press, Cambridge, MA 2007).

11. J. H. Myer. Optigami - A Tool for Optical System Design. Appl. Opt. 8, 260 (1969).

12. E. Cerda, L. Mahadevan. Confined developable elastic surfaces: cylinders, cones and the Elastica. Proc. R. Soc. A 461, 671–700 (2005).

13. A. H. Slocum. Precision Machine Design (Prentice Hall: Englewood Cliffs, NJ, 1992).

14. V. Doushkina, E. Fleming. Optical and mechanical design advantages using polymer optics. Advances in Optomechanics 7424 (2009)

15. C.-C. Lee, S.-Y. Hsiao, W. Fang. Formation and integration of a ball lens utilizing two-phase liquid technology. 2009 IEEE 22nd International Conference on Micro Electro Mechanical Systems 172-175 (2009).

16. K.-H. Jeong, G. L. Liu, N. Chronis, L. P. Lee. Tunable micro doublet lens array. Optics Express 31, 12:11 (2004).

17. O. Goldberg. Köhler illumination. The Microscope, 28, 15-22 (1980).

18. C. A. Cox, J. R. Carpenter. Improving attitudes toward teaching science and reducing science anxiety through increasing confidence in science ability in inservice elementary school teachers, Journal of Elementary Science Education, 1(2), 14-34 (1989).

19. T. Bredderman. The effects of activity-based science on student outcomes: A quantitative synthesis. Review of Education Research, 53, 499-518 (1983).

20. J. Bower. M. Scientists and science education reform: myths, methods, and madness. The National Academies RISE (Resources for Involving Scientists in Education), (1996).

21. M. Kremer, C. Brannen, R. Glennerster. The Challenge of Education and Learning in the Developing World, Science 297-300 (2013).

22. http://news.discovery.com/space/astronomy/supernova-discovered-by-10-year-old.htm

23. L. S. Garcia. Diagnostic medical parasitology, 5th ed., Washington, DC: ASM Press, 142–180 (2007).

24. E. J. Baron, et al. A Guide to Utilization of the Microbiology Laboratory for Diagnosis of Infectious Diseases: 2013 Recommendations by the Infectious Diseases Society of America (IDSA) and the American Society for Microbiology. Clinical Infectious Diseases, 57 (4): e22-e121 (2013).

25. A. W. Martinez, et al. Diagnostics for the Developing World: Microfluidic Paper-Based Analytical Devices. Anal. Chem. 82, 3-10 (2010).

26. K. Kikuchi, T. Morikawa, J. Shimada, K. Sakurai, Cladded radially inhomogeneous sphere lenses. Applied Optics 20, 388-394 (1981).

27. Y. Koike, A. Kanemitsu, Y. Shioda, E. Nihei, Y. Ohtsuka. Spherical gradient-index polymer lens with low spherical aberration. Applied Optics 33, 3394-3400 (1994).

**Acknowledgments:** We thank all members of the Prakash lab for valuable suggestions. Manu Prakash acknowledges support from Terman Fellowship, The Baxter Foundation, Coulter Foundation, Spectrum Foundation, C-Idea (National Institutes of Health grant RC4 TW008781-01), Bill and Melinda Gates Foundation, Pew Foundation, and Gordon and Betty Moore foundation for financial support. Jim Cybulski is supported by NIH Fogarty Institute Global Health Equity Scholars (GHES) Fellowship. James Clements was supported by NSF Graduate fellowship. We acknowledge Ioana Urama for assistance with supplementary figure S6, Anika Radiya for assistance with magnetic couplers and Marisa Borja for taking images in figure 5H-J, and Center for Discovery and Innovation in Parasitic Diseases (CDIPD) at UCSF and the Michael Hsieh Lab at Stanford University for supplying samples.

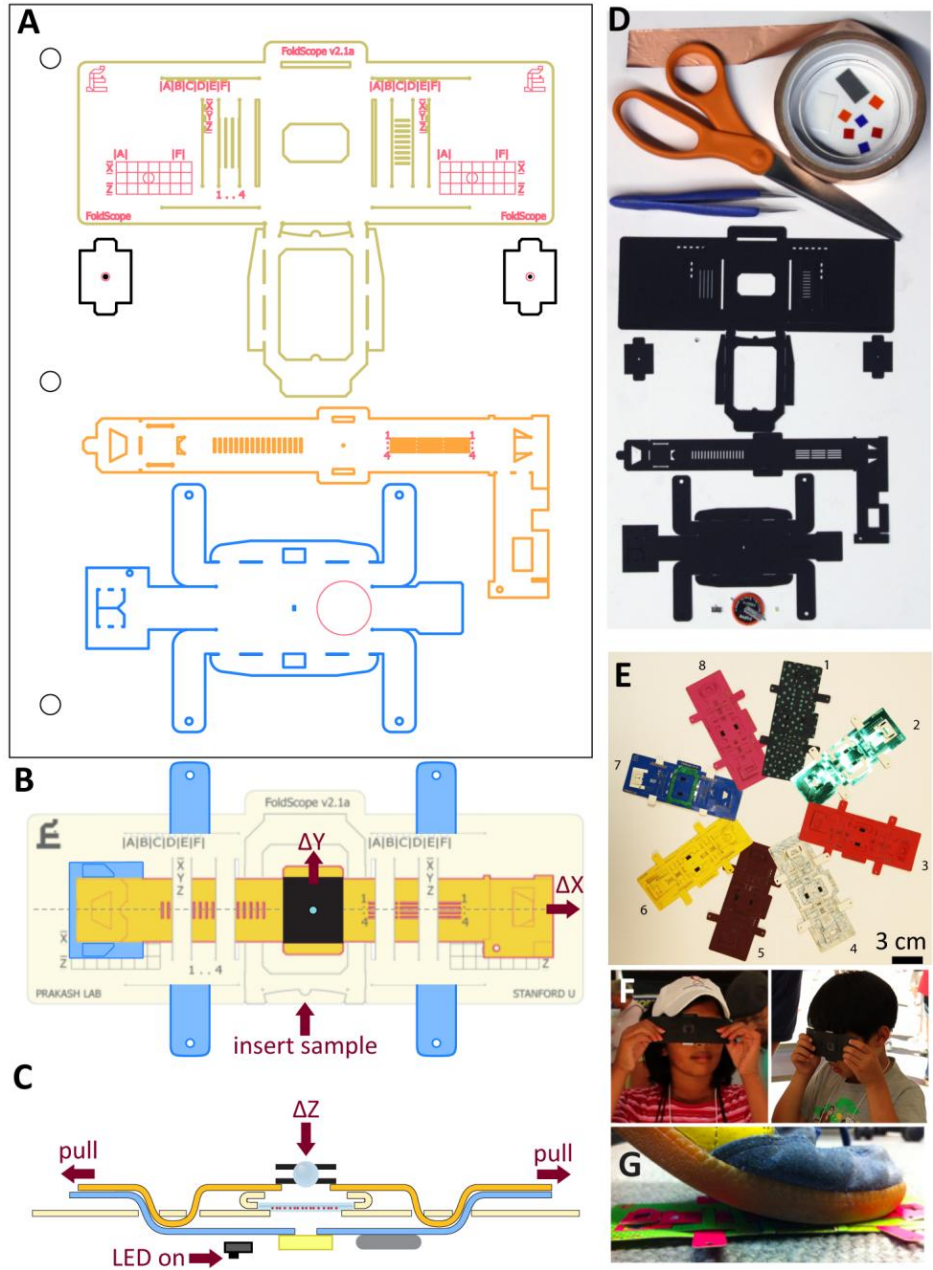

**Figure 1. Foldscope design, components and usage**. (A) CAD layout of Foldscope paper components on an A4 sheet. (B) Schematic of an assembled Foldscope illustrating panning, and (C) cross-sectional view illustrating flexure-based focusing. (D) Foldscope components and tools used in the assembly, including Foldscope paper components, ball lens, button-cell battery, surface-mounted LED, switch, copper tape and polymeric filters. (E) Different modalities assembled from colored paper stock. (F) Novice users demonstrating the technique for using the Foldscope. (G) Demonstration of the field-rugged design, such as stomping under foot.

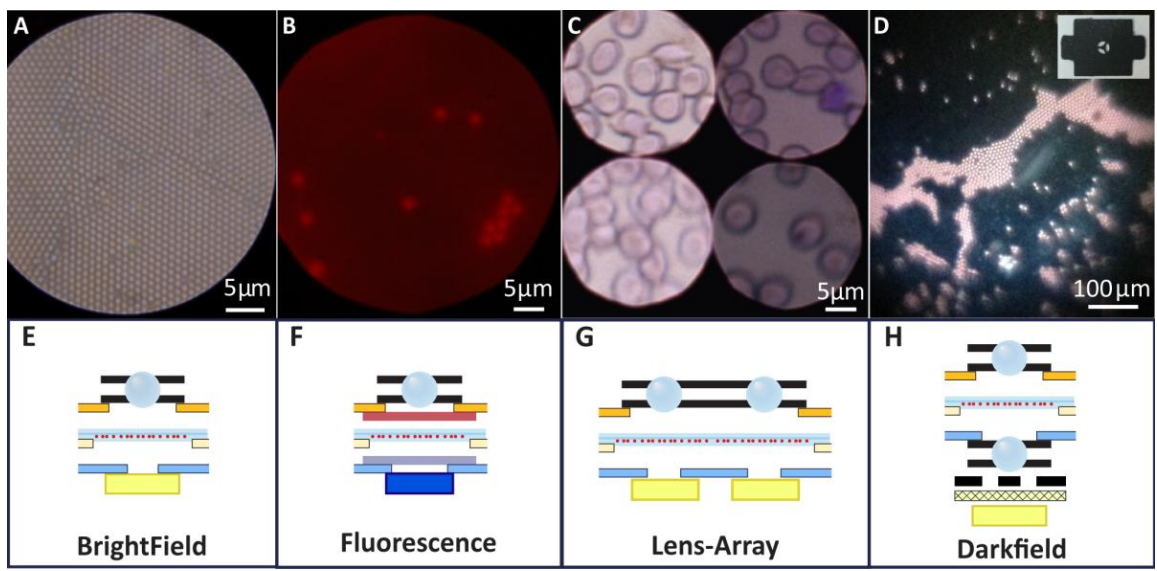

**Figure 2. Foldscope imaging modalities.** (A) Brightfield Foldscope image of a monolayer of 1μm polystyrene microspheres (Polysciences 07310-15) using a 1,450X lens. (B) Fluorescent Foldscope image of 2μm polyfluorescent microspheres (Polysciences 19508-2) using a 1,140X lens with Roscolux gel filters #19 and #80. (C) 2X2 lens-array Brightfield Foldscope image of Giemsa-stained thin blood smear using 1,450X lenses. (D) 140X Darkfield Foldscope images of 6μm polystyrene microspheres (Polysciences 15714-5), using a 140X lens for the darkfield condenser. Darkfield condenser aperture shown in inset has 1.5mm inner diameter and 4.0mm outer diameter. (E-H) Schematic cross-sections of Brightfield, Fluorescence, Lens-Array, and Darkfield Foldscope configurations, showing the respective arrangements of ball lenses, filters, and LEDs. See table 2 for ball lenses used for specific magnifications.

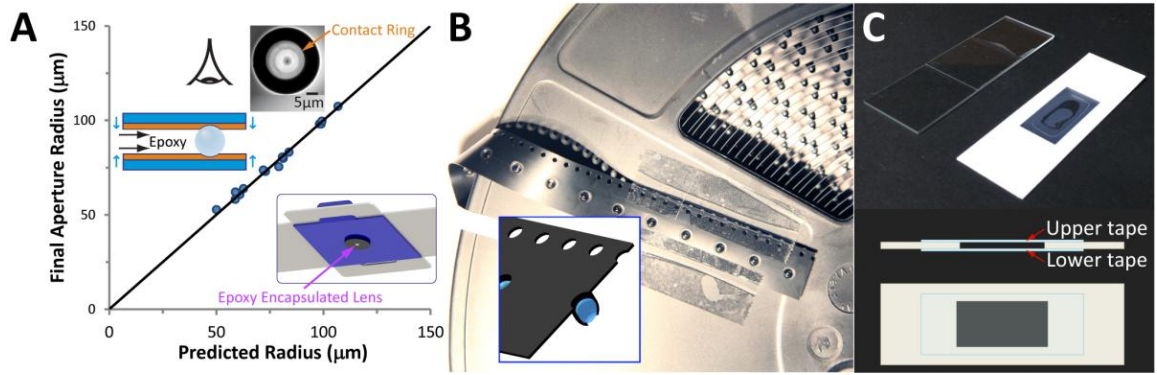

**Figure 3. Manufacturing innovations for lens- and specimen- mounting.** (A) Fabrication, mounting, and characterization of capillary-encapsulation process for lens-mounted apertures. X and Y error bars for all measurements are 2.5µm. (B) Reel of polystyrene carrier tape with custom pockets and punched holes for mounting over 2,000 ball lenses with optimal apertures. The first ten pockets include mounted ball lenses. Inset shows sectioned view from CAD model of carrier tape mounted lenses. Note the aperture is the punched hole shown on the bottom side of the ball lens. This tape is 16mm wide and is designed for 2.4mm ball lenses (aperture diameter is 0.7mm). (C) *Top:* Paper microscope slide shown next to standard glass slide with coverslip, both with wet mount algae specimens. *Bottom:* Schematic of paper microscope slide, showing specimen containment cavity formed between upper tape and lower tape in middle of slide.

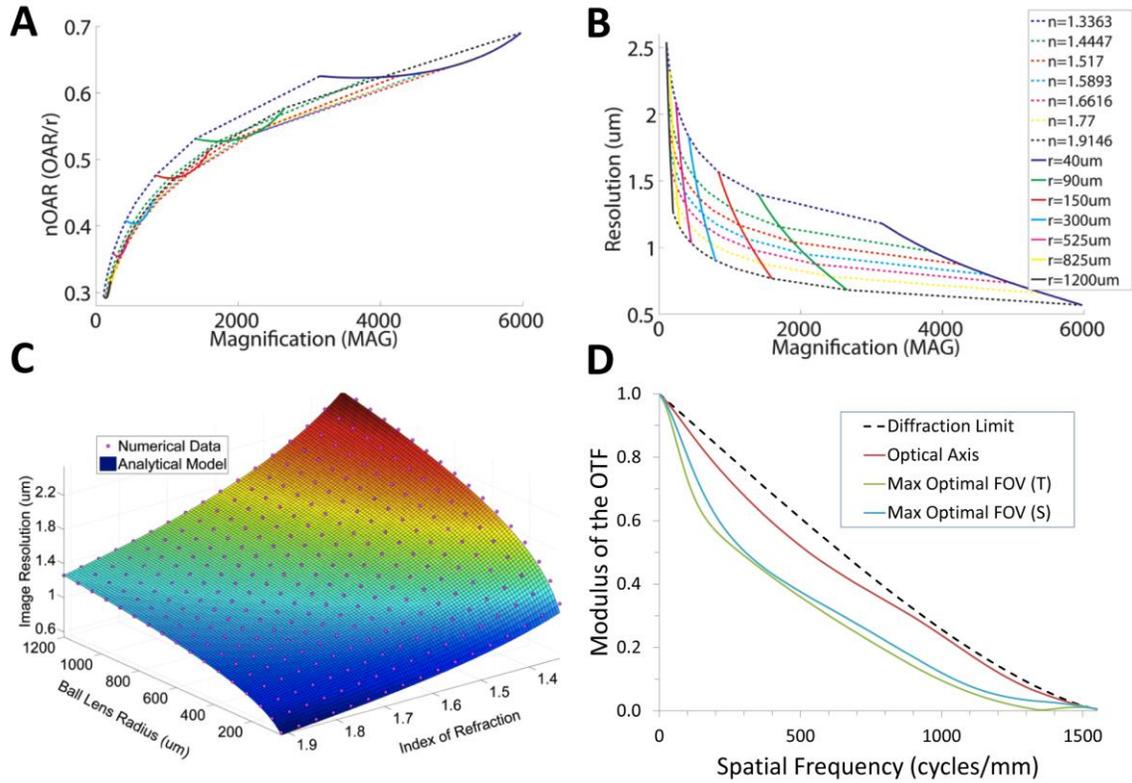

**Figure 4. Analytical and numerical modeling of Foldscope.** (A,B) Analytical "design curves" for normalized optimal aperture radius (nOAR) and optimal resolution (RES) versus magnification (MAG) over index of refraction (range 1.33-1.91) and ball lens radius (range 40-1200μm). (C) Comparision of analytical (3D surface) and numerical (plotted as points) results for RES versus index of refraction and ball lens radius. (D) Modulus of the Optical Transfer Function (MTF) over the optimal field of view for a 300μm sapphire lens with optimal aperture.

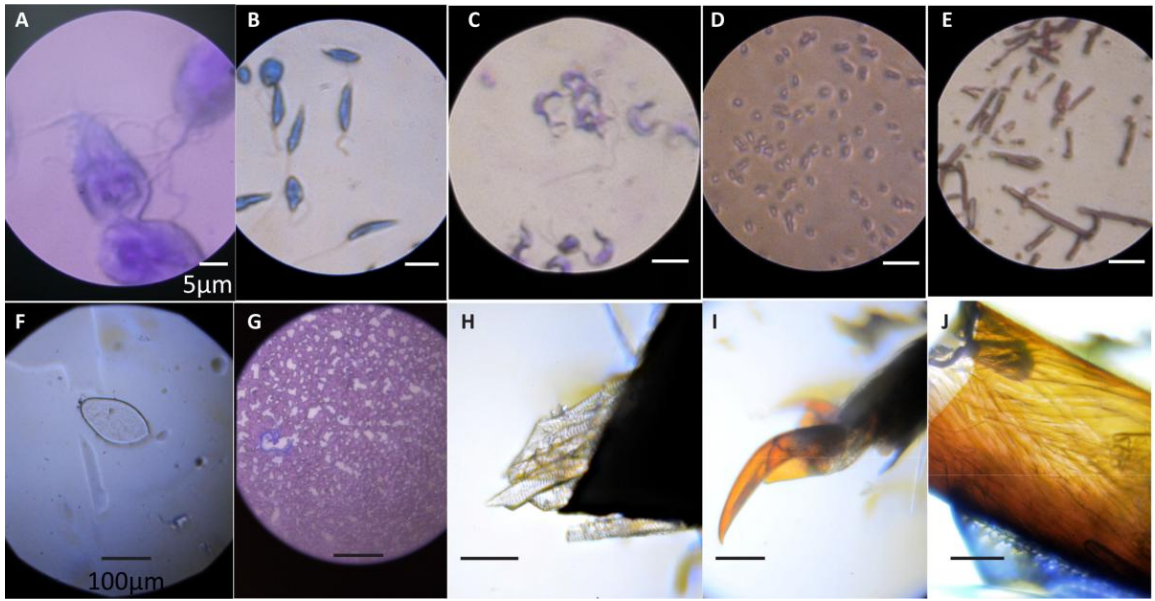

**Figure 5. Mosaic of Foldscope Images.** Bright field images of (A) *Giardia lamblia* (2,180X), (B) *Leishmania donovani* (1,450X), (C) *Trypanosom Cruzi* (1,450X), (D) gram-negative *Escherichia Coli* (1,450X), (E) gram-positive *Bacillus cereus* (1,450X), (F) *Schistosoma haematobium* (140X), and (G) *Dirofilaria immitis* (140X). Unstained (H) leg muscles and (I) tarsi of an unidentified ladybug (genus *Coccinella*). (J) Unstained leg muscles (fixed in formaldehyde) of an unidentified red ant (genus *Solenopsis*). An LED diffuser (Roscolux #111) was added for (A) and an LED condenser (2.4mm borosilicate ball lens) was used for (C). Images (H-J) were taken by novice user using a self-made Foldscope (140X). See table 2 for ball lenses used for specific magnifications. White scale bar: 5μm; black scale bar: 100μm.

**Table 1. Bill of Materials.**

| Component | Unit cost (10,000 pcs) |
|---|---|
| Paper (400 cm$^2$) | $0.06 |
| Ball Lens (low mag / high mag) | $0.17 / $0.56 |
| 3V button battery (CR2016) | $0.06 |
| LED | $0.21 |
| Switch | $0.05 |
| Copper Tape (5 cm$^2$) | $0.03 |
| Foldscope | $0.58 / $0.97 |

Summary of unit costs for Foldscope components in volumes of 10,000 units, not including assembly costs. This assumes a Foldscope in brightfield constructed from the following: polypropylene sheets (Press Sense 10mil Duropro); a 140X low-mag lens (Winsted Precision Ball 2.4mm borosilicate ball, P/N 3200940F1ZZ00A0) or a 2,180X high-mag lens (Swiss Jewel Co. 0.2mm sapphire ball lens); a 3V CR2016 button cell (Sanyo CR2016-TT1B #8565 from Batteriesandbutter.com); a white LED (Avago ASMT CW40 from Mouser.com); an electrical slider switch ("Off/On MINI SMD Switch" from AliExpress.com); and copper tape (Sparkfun P/N 76555A648).

**Table 2. Foldscope Analytical Model Parameter Summary Table.**

| Parameter | Functional Form in Optimized System | Parameter Values at Select Combinations of r,n | | | | | |
|---|---|---|---|---|---|---|---|
| | | r=1200 n=1.517 | r=500 n=1.517 | r=400 n=1.517 | r=150 n=1.517 | r=150 n=1.77 | r=100 n=1.77 |
| MAG | $MAG = \frac{2.5e5\,\mu m}{EFL} = (5e5\,\mu m) \cdot \frac{n-1}{n \cdot r}$ | 140 | 340 | 430 | 1140 | 1450 | 2180 |
| BFL | $BFL = EFL - r = \frac{1}{2} \cdot \frac{r \cdot (2-n)}{n-1}$ | 561 | 234 | 187 | 70 | 22 | 15 |
| RES | $RES = k_2 \cdot f \cdot (\lambda^3 \cdot |s|)^{1/4} = k_2 \cdot \left( \frac{\lambda^3 \cdot r \cdot n \cdot [n+(2-n) \cdot (2n-1)]}{128(n-1)^3} \right)^{1/4}$ | 1.90 | 1.52 | 1.44 | 1.13 | 0.86 | 0.77 |
| nOAR | $nOAR = \frac{a}{r} = \frac{k_1}{r} \cdot \left( \frac{\lambda}{|s|} \right)^{1/4} = k_1 \cdot \left( \frac{8\lambda \cdot n^3}{r \cdot (n-1) \cdot [n+(2-n) \cdot (2n-1)]} \right)^{1/4}$ | 0.294 | 0.366 | 0.387 | 0.495 | 0.510 | 0.565 |
| OAR | $OAR = nOAR \cdot r = k_1 \cdot \left( \frac{\lambda}{|s|} \right)^{1/4} = k_1 \cdot \left( \frac{8\lambda \cdot n^3 \cdot r^3}{(n-1) \cdot [n+(2-n) \cdot (2n-1)]} \right)^{1/4}$ | 353 | 183 | 155 | 74 | 77 | 56 |
| EFL | $EFL = \frac{1}{2} \cdot \frac{n \cdot r}{(n-1)}$ | 1761 | 734 | 587 | 220 | 172 | 115 |
| NA | $NA = \frac{2a \cdot (n-1)}{r \cdot n} = k_1 \cdot \left( \frac{128\,\lambda \cdot (n-1)^3}{r \cdot n \cdot [n+(2-n) \cdot (2n-1)]} \right)^{1/4}$ | 0.200 | 0.249 | 0.264 | 0.337 | 0.444 | 0.491 |
| FOV | $FOV = \frac{n \cdot a}{2(n-1)} = k_1 \cdot \left( \frac{\lambda \cdot r^3 \cdot n^7}{2(n-1)^5 \cdot [n+(2-n) \cdot (2n-1)]} \right)^{1/4}$ | 518 | 268 | 227 | 109 | 88 | 65 |
| DOF | $DOF = \frac{\lambda}{NA^2} = \frac{1}{k_1^2} \cdot \sqrt{\frac{\lambda \cdot r \cdot n \cdot [n+(2-n) \cdot (2n-1)]}{128(n-1)^3}}$ | 13.7 | 8.8 | 7.9 | 4.8 | 2.8 | 2.3 |
| SR | $SR = e^{-1/8}$ | 0.8825 | 0.8825 | 0.8825 | 0.8825 | 0.8825 | 0.8825 |

Functional form and select numerical values for the following dependent parameters: Magnification (MAG), Back Focal Length (BFL), Resolution (RES), nOAR (Normalized Optimal Aperture Radius), OAR (Optimal Aperture Radius), Effective Focal Length (EFL), Numerical Aperture (NA), Field of View (FOV), Depth of Field (DOF), Strehl Ratio (SR). These are calculated for infinite object distance per analytical model RM2, with aperture radius $a = OAR = k_1 \cdot (\lambda/s)^{1/4}$, $k_1 \cong 0.9321$, $k_2 \cong 0.7415$, and with aberration coefficient $s = -(n-1) \cdot [n+(2-n) \cdot (2n-1)]/(2r \cdot n)^3$.

# Supporting Information

**Supplementary Materials included:**
- Basic Expressions for Ball Lenses
- Analytical Model for an Foldscope in Brightfield
- Figures S1-S6
- Foldscope Assembly Video 1 (available online)
- Foldscope Drop Test Video 2 (available online)

**Supplementary Materials:**

**Basic Expressions for Ball Lenses.**

Using geometrical ray-tracing methods (see Figure S5B), the following expressions for optical properties of a ball lens can be derived under the paraxial approximation ($\sin\theta=\theta$):

$EFL = nr/2(n-1)$ (Effective focal length)
$BFL = (2-n)r/2(n-1)$ (Back focal length)
$MAG = 250mm/EFL$ (Magnification)
$FOV = na/2(n-1)$ (Field of View Radius)
$NA = 2a(n-1)/nr$ (Numerical Aperture)
$DOF = \lambda/NA^2$ (Depth of Field)

These expressions are written in terms of the following parameters: ball radius (r), index of refraction (n), aperture radius (a), and incident wavelength ($\lambda$). See Table S3 for the functional forms of these expressions for an optimized aperture and for evaluation of the expressions at select parameter values.

**Analytical Model for an Foldscope in Brightfield.**

*Introduction.* The primary parameters for the Foldscope optical system are object-lens distance, lens-image distance, ball lens radius, aperture radius, ball lens index of refraction, and incident illumination wavelength. The goal of this optimization is to determine the aperture radius and object-lens distance that will provide the smallest resolvable feature size, or in other words, that will minimize resolution. The "object plane" and "image plane" are interchanged relative to the physical system so that the size of the "image" in the model corresponds to the size of the object in the physical system. This technique is physically valid since time reversal symmetry applies to optical systems. It is useful since the time-reversed system naturally lends itself to computation of the smallest achievable spot size at the focal point in the "image plane", which corresponds to the smallest resolvable object (i.e., the resolution) in the physical system.

In the description that follows, note that the terms "image" and "object" refer to their respective entities in the time-reversed system. Also note that the sign convention for the numerical and analytical models is that the object-lens distance is negative for real objects, and the lens-image distance is positive for real images. In both cases, the sign is reversed if the object or image is virtual.

In this analysis, the system is modeled for the special case of an object at infinity, which physically corresponds to collimated light emerging from the Foldscope. The models are used to obtain functional relationships for the aperture radius and resolution (as well as other calculated quantities) in terms the physical parameters of the ball lens (radius and index of refraction) and of the incident illumination (wavelength).

This model effectively optimizes image resolution based on one field point at the center of the field of view. The advantage of this approach is that the resolution in this region will be the best achievable resolution for a simple ball lens, providing a critical capability for some applications. The disadvantage is that the edges of the field of view will have significant defocus. A numerical model was used to evaluate the image quality over the whole field, to define an "Optimal field of view" with good image quality, and to suggest strategies for extending the region of optimal resolution (see figure S4 and discussion in Design Innovations section of the main text).

Two analytical models were developed — one for each of the two resolution metrics, RM1 and RM2. These models predict the same exact functional forms for OAR and RES to a multiplicative constant, and these forms also show excellent agreement with the data from numerical modeling. The solution for these models is obtained in three steps. First, analytical expressions for OAR and RES are obtained in terms of the aberration coefficient of primary spherical aberration (s) for both resolution metrics. Second, an analytical expression is derived for the aberration coefficient, $s = s(n,r)$. Finally, these results are combined with an approximate expression for focal length $f = f(n,r)$ to yield the desired expressions, $OAR = OAR(\lambda,n,r)$ and $RES = RES(\lambda,n,r)$.

*Expressions from First Resolution Metric (RM1).* The first resolution metric is the absolute difference between the Airy Disk Radius (ADR) and RMS Spot Size (RSS),

$$RM1 \equiv |ADR - RSS| \qquad \text{(Eq. 1)}$$

This expression is minimized when,

$$ADR = RSS \qquad \text{(Eq. 2)}$$

The Airy Disk Radius is defined as (1),

$$ADR = 1.22 \cdot \lambda \cdot F \qquad \text{(Eq. 3)}$$

where F = the F/# or focal ratio, which is defined in terms of the focal length (f) and the aperture radius (a) as,

$$F = \frac{f}{2 \cdot a} \tag{Eq. 4}$$

RMS Spot Size can be approximated by RMS blur radius ($r_{RMS}$), which is given by (2,3),

$$RSS \cong r_{RMS} = 8F \cdot |S| \sqrt{\frac{1}{4} - \frac{\Lambda}{3} + \frac{\Lambda^2}{8}} \tag{Eq. 5}$$

where F = focal ratio, S = peak aberration coefficient, and $\Lambda$ = normalized longitudinal aberration. At best focus, $\Lambda=1$ and the approximate expression for RMS Spot Size simplifies to,

$$RSS \cong \frac{4}{\sqrt{6}} F \cdot |S| \tag{Eq. 6}$$

The peak aberration coefficient is given by,

$$S = s \cdot a^4 \tag{Eq. 7}$$

where s = the aberration coefficient of primary spherical aberration. The normalized optimal aperture radius (nOAR=OAR/r) is found by substituting into (Eq. 2) from (Eq. 3), (Eq. 6), and (Eq. 7) and solving for a/r. This yields,

$$nOAR_{RM1} = \frac{k_{1,RM1}}{r} \cdot \left(\frac{\lambda}{|s|}\right)^{1/4}, \qquad k_{1,RM1} = \left(\frac{1.22 \cdot \sqrt{6}}{4}\right)^{1/4} \cong 0.9297 \tag{Eq. 8}$$

The corresponding resolution is found by substituting a=OAR into the expression for Airy Disk Radius (or into the expression for RMS Spot Size). Substituting from (Eq. 4) and (Eq. 8) into (Eq. 3), the resolution for the first resolution metric is,

$$RES_{RM1} = k_{2,RM1} \cdot f \cdot \left(\lambda^3 \cdot |s|\right)^{1/4}, \qquad k_{2,RM1} = \left(\frac{1.22^3}{4\sqrt{6}}\right)^{1/4} \cong 0.6561 \tag{Eq. 9}$$

*Expressions from Second Resolution Metric (RM2).* As stated previously, the second resolution metric is the Airy Disk Radius (ADR) divided by the Strehl Ratio (SR). This is minimized when its derivative with respect to the aperture radius equals zero,

$$\frac{\partial}{\partial a}\left(\frac{ADR}{SR}\right) = 0 \tag{Eq. 10}$$

Applying the quotient rule leads to the following equivalent condition:

$$SR \cdot \frac{\partial ADR}{\partial a} = ADR \cdot \frac{\partial SR}{\partial a} \tag{Eq. 11}$$

The Strehl Ratio is approximated by a simple empirical expression given by Mahajan (4),

$$SR = e^{-(2\pi\omega_s/\lambda)^2} \tag{Eq. 12}$$

where $\omega_s$ = RMS wavefront error at the best focus. The RMS wavefront error due to spherical aberration at best focus is given by (2,3),

$$\omega_s = \frac{S}{\sqrt{180}} \tag{Eq. 13}$$

where S = peak aberration coefficient as given by (Eq. 7). Substituting (Eq. 7) and (Eq. 13) into (Eq. 12), the Strehl ratio can be written as,

$$SR = e^{-C_1 a^8}, \qquad C_1 = \left(\frac{\pi \cdot s}{3\sqrt{5} \cdot \lambda}\right)^2 \tag{Eq. 14}$$

Using (Eq. 3) and (Eq. 4), the Airy Disk Radius can be written as,

$$ADR = \frac{C_2}{a}, \qquad C_2 = \frac{1.22 \cdot \lambda \cdot f}{2} \tag{Eq. 15}$$

The expression for the second resolution metric is therefore,

$$\frac{ADR}{SR} = \frac{C_2}{a} \cdot e^{C_1 a^8} \tag{Eq. 16}$$

Taking the derivatives of (Eq. 14) and (Eq. 15) with respect to the aperture radius yields,

$$\frac{\partial SR}{\partial a} = -8 C_1 \cdot a^7 \cdot e^{-C_1 a^8} \tag{Eq. 17}$$

$$\frac{\partial ADR}{\partial a} = \frac{-C_2}{a^2} \tag{Eq. 18}$$

Substituting (Eq. 14), (Eq. 15), (Eq. 17), and (Eq. 18) into (Eq. 11) and solving for a/r gives the following expression for the normalized optimal aperture radius (nOAR=OAR/r) corresponding to the second resolution metric,

$$nOAR_{RM2} = \frac{k_{1,RM2}}{r} \cdot \left(\frac{\lambda}{|s|}\right)^{1/4}, \qquad k_{1,RM2} = \left(\frac{3\sqrt{10}}{4\pi}\right)^{1/4} \cong 0.9321 \tag{Eq. 19}$$

The corresponding resolution is found by substituting a=OAR into the expression for ADR/SR. Substituting (Eq. 19) into (Eq. 16), and eliminating $C_1$ and $C_2$ using (Eq. 14) and (Eq. 15), the resolution for the second resolution metric is,

$$RES_{RM2} = k_{2,RM2} \cdot f \cdot \left(\lambda^3 \cdot |s|\right)^{1/4}, \qquad k_{2,RM2} = 1.22 \cdot \left(\frac{\pi}{12}\right)^{1/4} \cdot \left(\frac{e}{10}\right)^{1/8} \cong 0.7415 \quad \text{(Eq. 20)}$$

*Comparison of RM1 Model to RM2 Model.* Note that the two resolution metrics RM1 and RM2 yield identical functional forms for nOAR and RES. Compare (Eq. 8) to (Eq. 19) and (Eq. 9) to (Eq. 20). Their predicted values for $k_1$ differ only by 0.26%, indicating excellent agreement for the size of the optimal aperture. Since RM2 is a more conservative metric that includes the Strehl Ratio, it is not surprising that it predicts coarser resolution, with $k_2$ 11.5% larger than the RM1 model.

Also note that the analyses of the two models are perfectly general up to here, within the limits of the approximations in (Eq. 6) for RMS Spot Size and (Eq. 12) for Strehl Ratio. Therefore, the expressions for nOAR and RES may be applied to other, more complex systems (not just ball lenses) to determine values for optimal aperture radius and resolution, given expressions for the aberration coefficient (s) and the focal length (f) for such a system.

*Expression for aberration coefficient for primary spherical aberration.* This system consists of two optical surfaces and has a circular aperture concentrically located on the first surface. To find an expression for the aberration coefficient ($s_i$) for the i[th] surface of a system, the wave aberration (W) is first calculated for that surface as the optical path difference between the chief ray and a marginal ray. The aberration coefficient (s) is then obtained as the coefficient of the term with aperture radius to the 4[th] power. From the derivation by Mahajan (4), the resulting expression for the aberration coefficient of the i[th] surface is,

$$s_i = -\frac{n_i'^2}{8} \cdot \left(\frac{1}{R_i} - \frac{1}{L_i'}\right)^2 \cdot \left(\frac{1}{n_i' L_i'} - \frac{1}{n_i L_i}\right) \qquad \text{(Eq. 21a)}$$

where $n_i$, $n_i'$ are the refractive indices of the media before and after the surface; $L_i$, $L_i'$ are the object-lens distance and the lens-image distance for the i[th] surface; and $R_i$ is the ball lens radius (positive if the arc is centered to the right of the surface, and negative if the arc is centered to the left of the surface). The following alternate forms,

$$s_i = -\frac{n_i' \cdot (n_i' - n_i)}{8 n_i^2} \cdot \left(\frac{1}{R_i} - \frac{1}{L_i'}\right)^2 \cdot \left(\frac{n_i'}{R_i} - \frac{n_i - n_i'}{L_i'}\right) \qquad \text{(Eq. 21b)}$$

$$s_i = -\frac{n_i^2 \cdot (n_i - n_i')}{8 n_i'^2} \cdot \left(\frac{1}{R_i} - \frac{1}{L_i}\right)^2 \cdot \left(\frac{n_i + n_i'}{n_i L_i} - \frac{1}{R_i}\right) \qquad \text{(Eq. 21c)}$$

can be obtained by eliminating $L_i$ and $L_i'$, respectively, from (Eq. 21a) using the Gaussian imaging equation (1),

$$\frac{n_i'}{L_i'} - \frac{n_i}{L_i} = \frac{n_i' - n_i}{R_i} \quad \text{(Eq. 22)}$$

Using subscripts to denote the surface, the indices of refraction of this system are $n_1 = n_2' = 1$ and $n_1' = n_2 = n$, where n = index of refraction of the ball lens. The radii of the respective surfaces are $R_1 = -R_2 = r$, where r = the radius of curvature of the ball lens. In this model, the object is assumed to be at infinity ($L_1$ = infinity), so $L_1'$ equals the focal length ($f_1$). Substituting the values for the first surface in (Eq. 22) and solving for the lens-image distance ($L_1'$) gives,

$$L_1' = f_1 = \frac{n \cdot r}{n-1} \quad \text{(Eq. 23)}$$

Since the image plane of the first surface serves as the object plane of the second surface, and the two surfaces are separated by twice the ball lens radius, the object-lens distance for the second surface can be written as,

$$L_2 = L_1' - 2r \quad \text{(Eq. 24)}$$

Substituting (Eq. 23) into (Eq. 24) gives,0

$$L_2 = \frac{r \cdot (2-n)}{n-1} \quad \text{(Eq. 25)}$$

Writing out (Eq. 21c) for i=1 and i=2, and substituting in the corresponding values for these surfaces gives,

$$s_1 = -\frac{n-1}{8n^2 r^3} \quad \text{(Eq. 26)}$$

$$s_2 = -\frac{n \cdot (n-1) \cdot (2n-1)}{8r^3 \cdot (2-n)^3} \quad \text{(Eq. 27)}$$

The total aberration coefficient is the sum of the aberration coefficients from the two surfaces, with the coefficient for the second surface weighted by the fourth power of the ratio of the effective aperture radius on the second surface (e) to the aperture radius on the first surface (a),

$$s = s_1 + \left(\frac{e}{a}\right)^4 \cdot s_2 \quad \text{(Eq. 28)}$$

From the ray tracing diagram in figure S5B, it is evident that the ratio e/a can be written as,

$$\frac{e}{a} = \frac{r \cdot \sin(2\beta - \alpha)}{r \cdot \sin(\alpha)} \tag{Eq. 29}$$

where α = angle of incidence of incoming collimated light, β = angle of incidence of light inside the glass, and r = radius of the ball lens. A relation between the angles α and β is provided by Snell's Law,

$$\sin(\alpha) = n \cdot \sin(\beta) \tag{Eq. 30}$$

Applying the paraxial approximation sinθ=θ to (Eq. 29) and (Eq. 30) and combining the results gives the following approximate expression for e/a,

$$\frac{e}{a} \cong \frac{2\beta}{\alpha} - 1 \cong \frac{2-n}{n} \tag{Eq. 31}$$

Substituting (Eq. 26), (Eq. 27), and (Eq. 31) in (Eq. 28) gives the following expression for the aberration coefficient,

$$s = -\frac{(n-1) \cdot [n + (2-n) \cdot (2n-1)]}{(2r \cdot n)^3} \tag{Eq. 32}$$

While this expression was derived independently as described above, it is equivalent to an expression for the Sidel coefficient $S_I$ for a thick lens recently published in Applied Optics by Miks and Novak (5).

*Expressions for OAR=OAR(λ,n,r) and RES=RES(λ,n,r).* Now that an expression has been obtained for the aberration coefficient s(n,r), all that is needed to find OAR(λ,n,r) and RES(λ,n,r) is an expression for the focal length f(n,r). This is given by the Lensmaker's equation (1),

$$\frac{1}{f} = (n-1) \cdot \left[\frac{1}{R_1} - \frac{1}{R_2} + \frac{(n-1) \cdot d}{n \cdot R_1 \cdot R_2}\right] \tag{Eq. 33}$$

where n = index of refraction of the ball lens; $R_1, R_2$ = radii of curvature of the respective surfaces; and d = spacing between the surfaces. Substituting $R_1$=-$R_2$=r and d=2r, the effective focal length (EFL) for a ball lens is,

$$f = EFL = \frac{r \cdot n}{2(n-1)} \tag{Eq. 34}$$

Since the resolution metrics RM1 and RM2 both yielded the same functional form for nOAR and RES in terms of s(n,r) and f(n,r), their final results will also be the same within a multiplicative constant. Substituting (Eq. 34) and (Eq. 32) into (Eq. 8) and (Eq. 9) for RM1 and into (Eq. 19) and (Eq. 20) for RM2, the final expressions for normalized optimal aperture radius (nOAR) and resolution (RES) are,

$$nOAR = k_1 \cdot \left( \frac{8\lambda \cdot n^3}{r \cdot (n-1) \cdot [n + (2-n) \cdot (2n-1)]} \right)^{1/4} \quad \text{(Eq. 35)}$$

$$RES = k_2 \cdot \left( \frac{\lambda^3 \cdot r \cdot n \cdot [n + (2-n) \cdot (2n-1)]}{128(n-1)^3} \right)^{1/4} \quad \text{(Eq. 36)}$$

with the following values for the multiplicative constants $k_1$ and $k_2$ as given by (Eq.8), (Eq.9), (Eq.19), and (Eq.20),

$$k_{1,RM1} = \left( \frac{1.22 \cdot \sqrt{6}}{4} \right)^{1/4} \cong 0.9297 \quad \text{(Eq. 37)}$$

$$k_{2,RM1} = \left( \frac{1.22^3}{4\sqrt{6}} \right)^{1/4} \cong 0.6561 \quad \text{(Eq. 38)}$$

$$k_{1,RM2} = \left( \frac{3\sqrt{10}}{4\pi} \right)^{1/4} \cong 0.9321 \quad \text{(Eq. 39)}$$

$$k_{2,RM2} = 1.22 \cdot \left( \frac{\pi}{12} \right)^{1/4} \cdot \left( \frac{e}{10} \right)^{1/8} \cong 0.7415 \quad \text{(Eq. 40)}$$

**Supplementary References:**

1. E. Hecht. Optics. 2nd ed., Addison Wesley (1987).
2. V. Sacek. Amateur Telescope Optics website, http://www.telescope-optics.net, (2006).
3. V. N. Mahajan. Aberration Theory Made Simple. 2nd ed., SPIE Press, Bellingham, Washington (2011).
4. V. N. Mahajan, Optical Imaging and Aberrations: Wave Diffraction Optics. *SPIE-International Society for Optical Engineering Press* (1998).
5. A. Miks, J. Novak. Third-order aberration coefficients of a thick lens. Applied Optics, 51(33): 7883-86 (2012).
6. http://www.theimage.com/digitalphoto2/index7b.html
7. M. Allen, L. Weintraub, et al. Forensic Vision With Application to Highway Safety, Lawyers and Judges Publishing Co, Inc, Tucson, Arizona, 467 (2008).

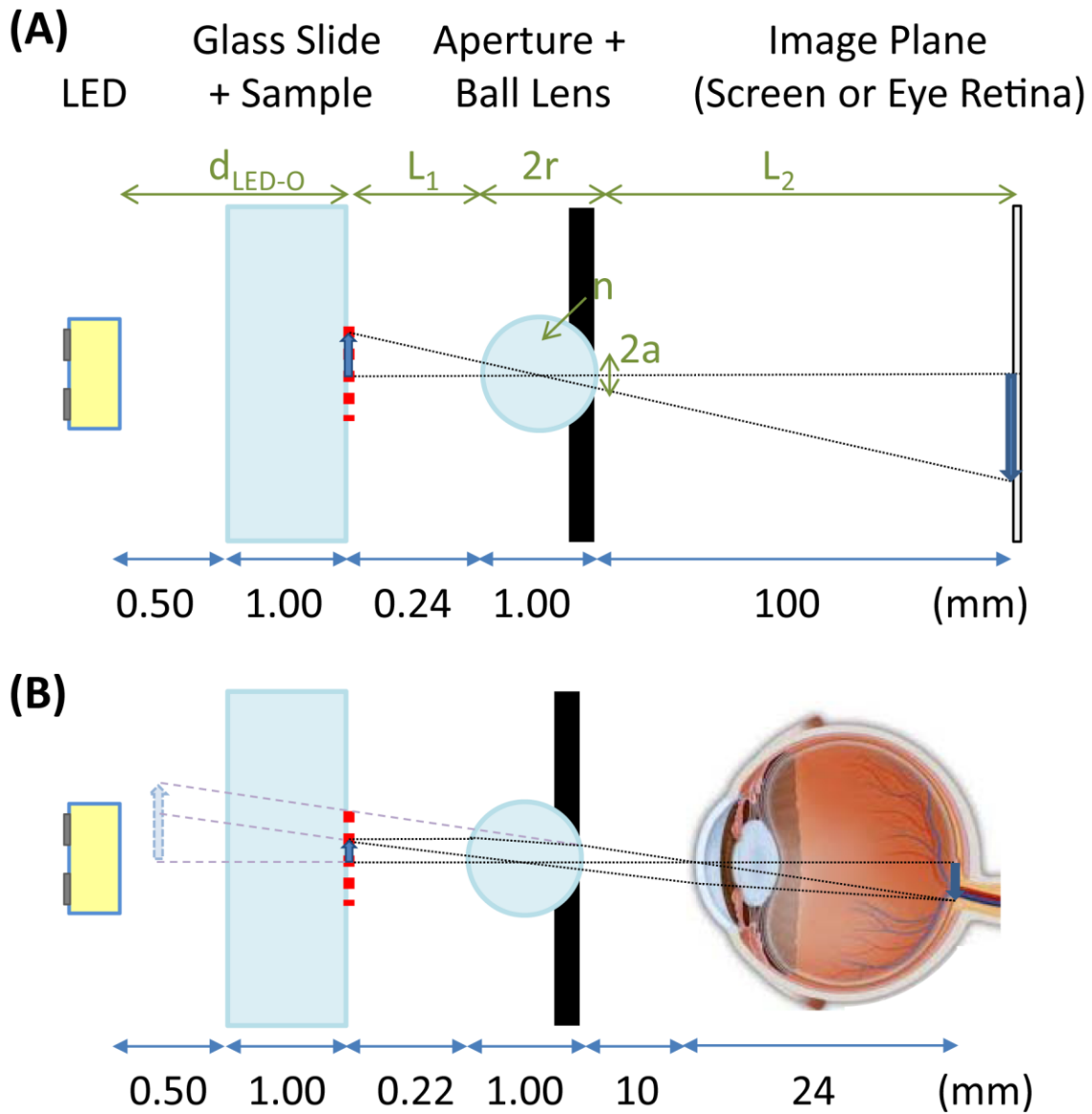

**Figure S1. Foldscope Schematics.** (A) Real image formation via projection. (B) Virtual image formation via direct observation with the eye. Note the drawings are not to scale. The indicated lengths are example values that show the versatility of this design as well as its extreme space efficiency. For example, the same system can be used for projecting or imaging simply by changing the object-lens distance by about 20 μm. Also, notice the total path length from the LED to the lens is almost an order of magnitude smaller than the size of the human eye.

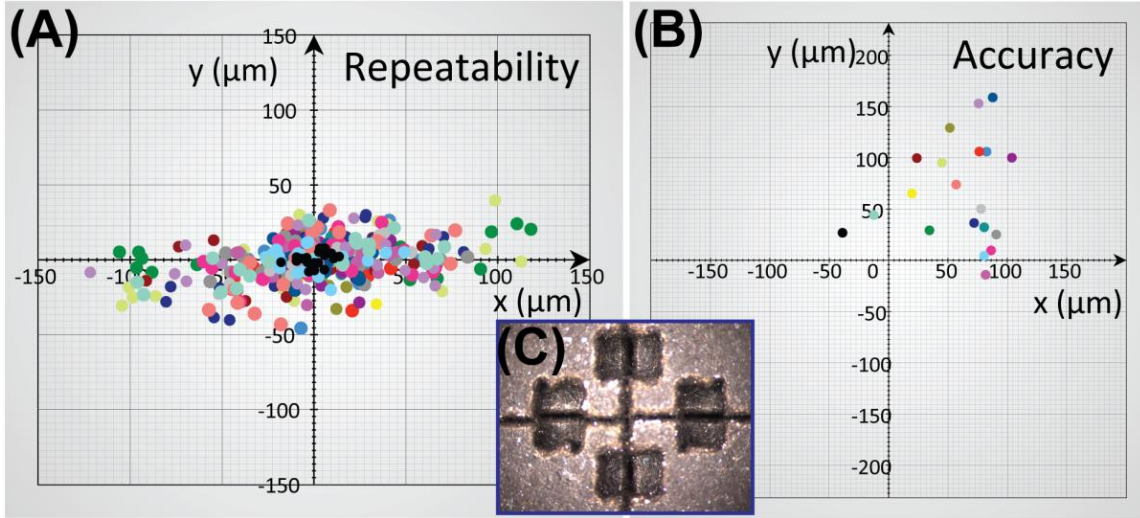

**Figure S2. Characterization of Self-Alignment by Folding.** Twenty independent Foldscopes were constructed out of 350μm thick black cardstock and manually folded and unfolded twenty times each, with alignment measured after each assembly. The data was used to produce plots of (A) assembly repeatability (distribution of all 400 values, adjusted to give zero mean for each Foldscope) and (B) assembly accuracy (distribution of 20 mean values calculated per Foldscope) using (C) cross-hair alignment features on the optics and illumination stages. Note that the span of the data in both plots is less than the thickness of the paper used to construct the Foldscopes. Based on the data shown in the plots, assembly repeatability was assessed as the mean value of twice the standard deviation for each Foldscope (65μm in X and 25μm in Y), while assembly accuracy was assessed as the mean value of all trials (59μm in X and 67μm in Y). A higher skew in X-axis repeatability results from structurally distinct constraints implemented for the X- and Y-axes, while the assembly accuracy errors in both directions are consequences of the design which can be compensated by feature shifts in future designs. Note that the X and Y error bars for all measurements are 8.4μm.

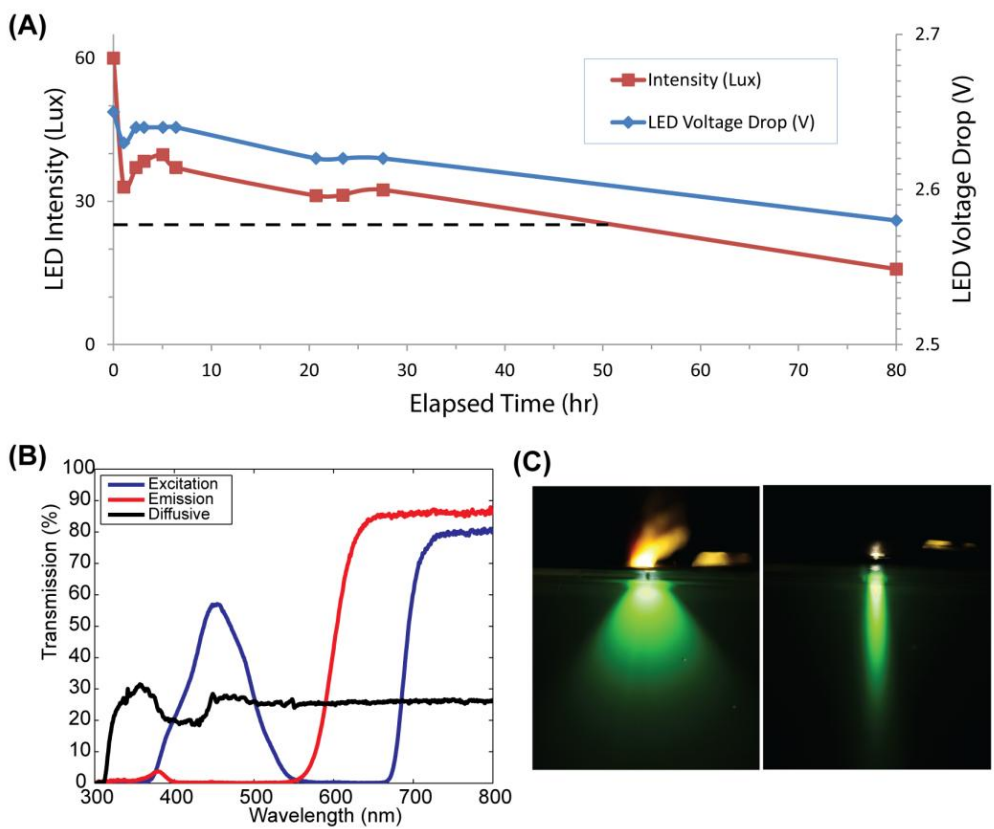

**Figure S3. Component Characterization.** (A) LED voltage and intensity versus time for a white LED (Avago HSMW-CL25) powered by a Duracell CR2032 battery with no resistor. (B) Filter transmission spectra of three Roscolux filters ─ Tough Rolex diffuser (#111), Fire Red (#19), and Primary Blue (#80) ─ measured with Ocean Optics Photo spectrometer USB4000. (C) Intensity profile of a white LED (Avago HSMW-CL25) as visualized in water with dissolved fluorescein. The left image is taken with the bare LED while the right image is taken with a condenser lens (2.4mm borosilicate ball lens) placed adjacent to the LED in the optical path, demonstrating that a ball lens can be used to effectively collimate the light emitted by this LED.

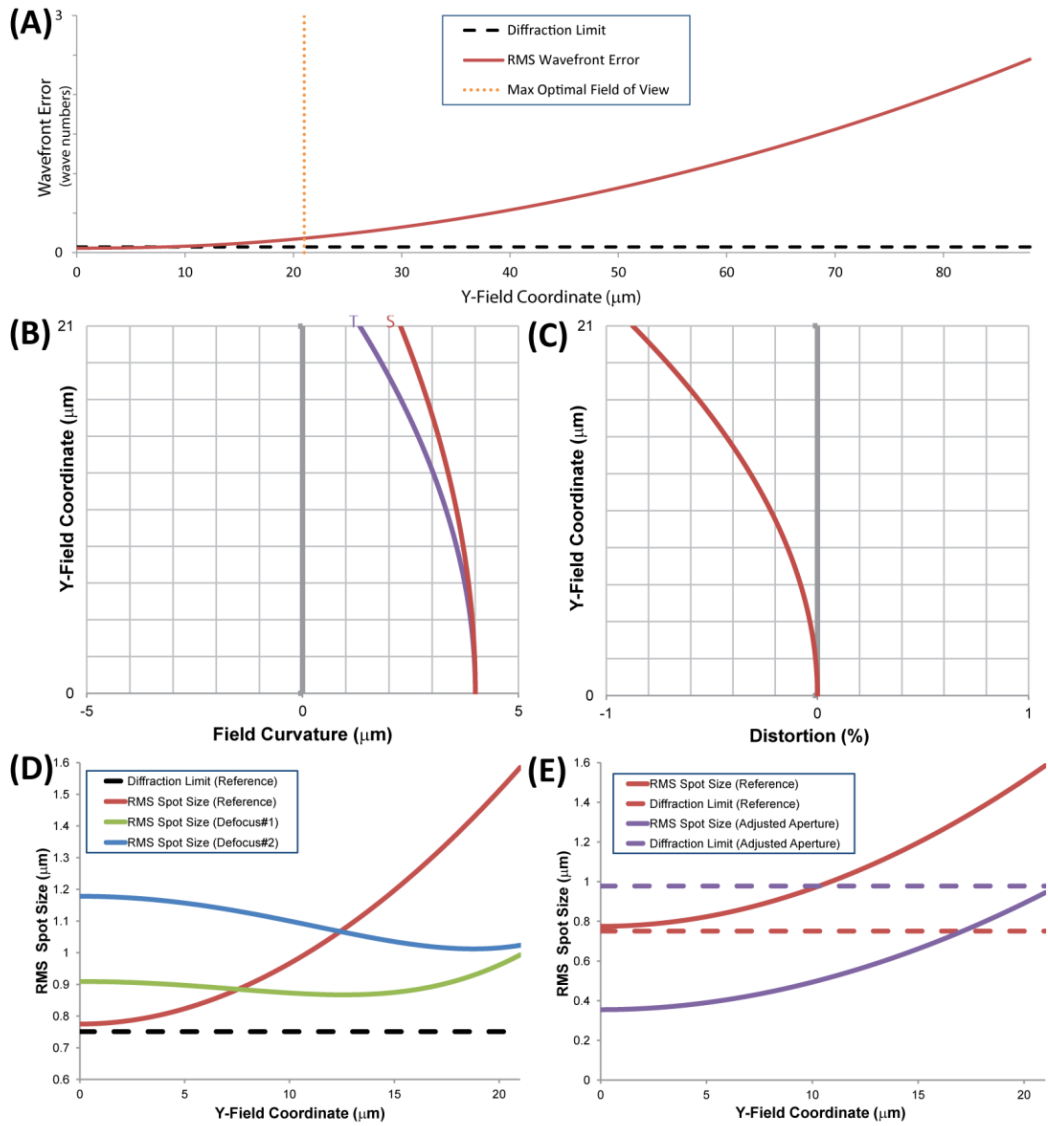

**Figure S4. Numerical Modeling Characterization of Optimal Field of View.** (A) Plot of Wavefront Error over the full field of view defined by the aperture for a 300μm Sapphire ball lens with a 147μm aperture. With increasing field coordinate, the Wavefront Error becomes very large and the image will be out of focus. An "optimal field of view" is defined at a field coordinate of 21μm, where the Wavefront Error is approximately 1/5 wave number. (B,C) Plots of Field Curvature and Distortion over the optimal field of view. (D) Plot of RMS spot size over the optimal field of view depicting four cases: optimized solution treated as reference with zero defocus (red line), defocus of 3μm (green line), defocus of 3μm (green line), diffraction limit (dashed black line). The reference solution provides the best achievable resolution at the center of the field of view (approximately equal to the diffraction limit for this choice of aperture), while other plots show that increasing defocus moves the region of best resolution radially out from the center in an annular ring. (E) Plot of RMS spot size over the optimal field of view depicting optimal aperture predicted by analytical model (red lines) and adjusted aperture giving uniform RMS spot size over the field of view (purple lines).

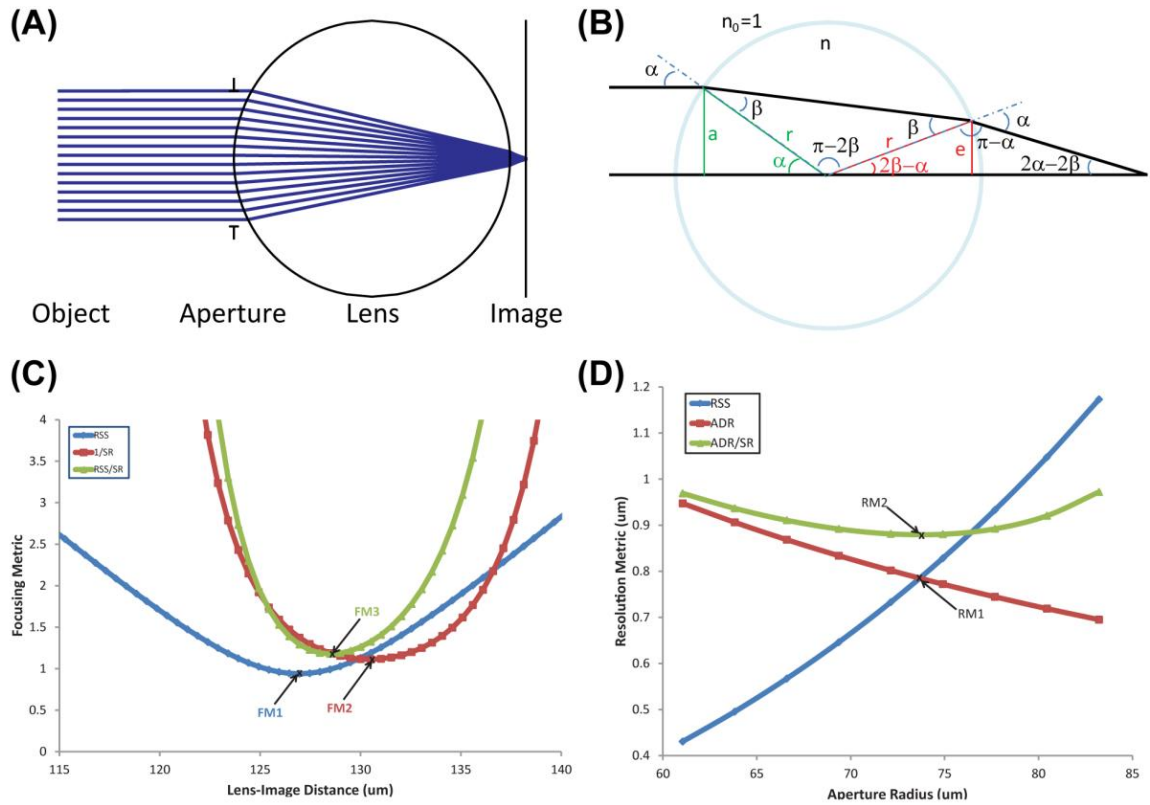

**Figure S5. Diagrams and Plots for Numerical and Analytical Models.** (A) Schematic of time-reversed Zemax model showing collimated light coming from an object at infinity, passing through aperture, and focused by the ball lens onto a focal point in the image plane. (B) Schematic of time-reversed model showing key parameters used in some derivations for the analytical model. (C) Plot of Focusing Metric versus Lens-Image Distance for $\lambda=0.55\mu m$, $r=150\mu m$, $n=1.517$. This illustrates how focusing metrics FM1, FM2, and FM3 select different values for the optimal lens-image distance. (D) Plot of Resolution Metric versus Aperture Radius for $\lambda=0.55\mu m$, $r=150\mu m$, $n=1.517$. This illustrates how resolution metrics RM1 and RM2 select nearly the same aperture radius but yield different values for resolution.

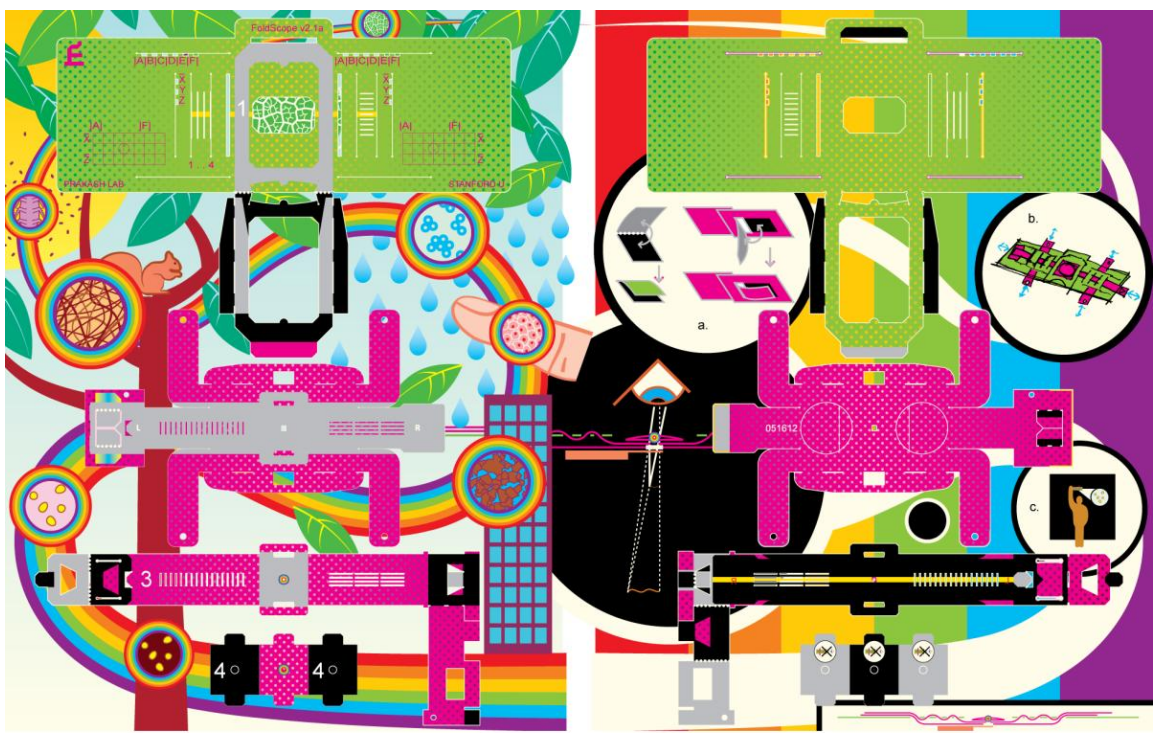

**Figure S6. Artistic Layout of Foldscope Paper Components.** Artistic version of Foldscope layout with integrated universal folding instructions based on color coding, where like colors are matched during the folding process to leave a single solid color in the final folded configuration.

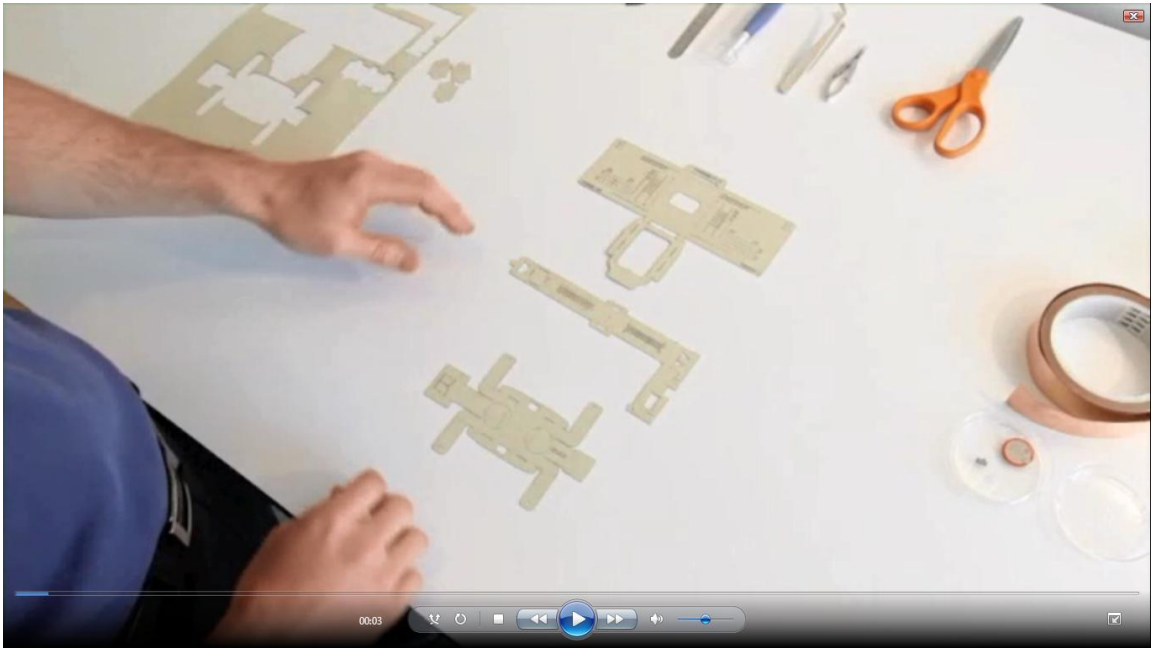

**Movie S1. Foldscope Assembly.** A short video of 140X Brightfield Foldscope assembly process.

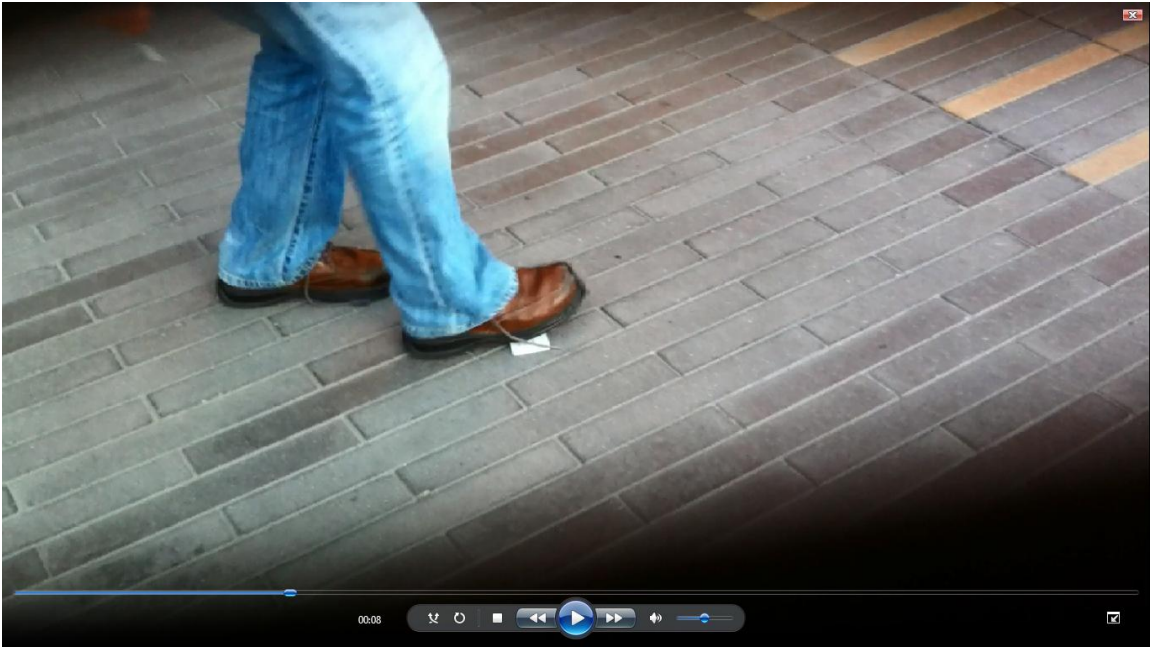

**Movie S2. Foldscope Drop Test and ruggedness.** A short video of a three story drop test and ruggedness of Foldscope demonstrated by stomping under feet.